\documentclass[12pt]{article}
\begin{document}

\vspace*{1.0 cm}

\begin{center}
{\Large \bf Toward an Infinite-component Field Theory with a Double Symmetry:
Free Fields} \\

\vspace{1.0 cm}

{\large L.M.Slad \footnote{E-mail: slad@theory.sinp.msu.ru}} \\

\vspace{0.4 cm}

{\it D.V.Skobeltsyn Institute of Nuclear Physics, \\
Moscow State University, Moscow 119899}
\end{center}

\vspace{0.5 cm}

\centerline{\bf Abstract}

\begin{small}
We begin a study of possibilities of describing hadrons in terms of monolocal 
fields which transform as proper Lorentz group representations decomposable into
an infinite direct sum of finite-dimensional irreducible representations. The
additional requirement that the free-field Lagrangians be invariant under the 
secondary symmetry transformations generated by the polar or the axial 
four-vector representation of the orthochronous Lorentz group provides an 
effective mechanism for selecting the class representations considered and 
eliminating an infinite number of arbitrary constants allowed by the 
relativistic invariance of the Lagrangians. 
\end{small}

\vspace{0.5 cm}

\begin{center}
{\large \bf 1. Introduction}
\end{center}

The first attempt, by Ginzburg and Tamm [1], at a relativistic description of 
particles with internal degrees of freedom was based on bilocal equations. It
turned out that each of the considered equations yielded a nonadmissible mass 
spectrum, namely, it was shown that the masses tended to zero as some 
state-specifying quantum number increased. To clear up the question of how 
general this result is, Gelfand and Yaglom provided a complete description of 
all linear relativistically invariant equations of the form
\begin{equation}
(\Gamma^{\mu} \partial_{\mu} + i R) \Psi (x) = 0,
\end{equation}
where $\Gamma^{\mu}$ and $R$ are matrix operators, and the corresponding 
Lagrangians [2]. Having restricted themselves to the consideration of 
equations (1), in which the fields $\Psi (x)$ transform as proper Lorentz 
group representations decomposable into a finite direct sum of 
infinite-dimensional irreducible representations (in what follows, we call 
such fields the FSIIR-class fields) and the operator $R$ is nondegenerate, 
Gelfand and Yaglom concluded that the masses tend to zero with infinitely 
increasing spin. It was later shown [3] that this conclusion failed in some 
cases: there are FSIIR-class representations of the proper Lorentz group and 
restrictions on arbitrary constants in Eq. (1) such that the masses tend to 
zero with increasing spin in some branches of the mass srectrum and tend to 
infinity in the other branches. It was also shown that if the matrix operator 
$R$ is degenerate (this case was omitted in [2]), there are equations of type 
(1) for the FSIIR-class fields such that the nonzero masses in the 
corresponding mass spectrum tend to infinity with increasing spin; however, in 
this case, there are massless states for all spins beginning with some minimum 
value $s_{0}$. These corrections to the Gelfand-Yaglom conclusion do not annul 
its physical content: the mass spectrum obtained in the framework of the 
Lagrangian approach for any of the FSIIR-class fields has peculiarities that 
are absolutely unacceptable in particle physics.

A rather general property of the FSIIR-class fields is their local 
noncommutativity, $[\Psi (x), \Psi (y)]_{\pm} \neq 0$ for $(x-y)^{2}<0$, proved
in [4] under the assumption that the set of field states is complete and the 
mass spectrum does not contain infinitely degenerate levels. In the opinition 
of the authors of [5], it seems likely that this difficulty requires weakening 
the strict locality postulate.

The connection between spin and statistics for an infinite-component field
(not necessarily of the FSIIR-class), satisfying Eq.(1), holds or does not hold
depending upon what irreducible representations of the proper Lorentz group
compose the representation which desribes the field under consideration [6].

Some of the revealed "deceases" of the FSIIR-class fields, namely, the 
existence of spacelike solutions of the Gelfand-Yaglom equations [7] and the 
lack of $CPT$-invariance [8], are not common to all of them (the respective
counterexamples can be found in [9] and [10]).
 
The result of all previous studies of the infinite-component fields can be 
summarized as follows.

First, the hopelessness of such a description of particles is established only
for one class of the infinite-component fieds, the FSIIR-class. The fields that 
transform as proper Lorentz group representations decomposable into an infinite
direct sum of finite-dimentional or infinite-dimentional irreducible 
representations (in what follows, we call these fields the respective 
ISFIR-class or ISIIR-class fields) have not been investigated so far. The 
reason for this is an infinite number of arbitrary constants in the 
relativistically invariant Lagrangians for such fields.

Second, having been convinced that the general properties of the 
finite-component fieds and the FSIIR-class fields are different in many 
respects, we can expect that the general properties of the FSIIR-class and
ISFIR-class fields are also different.

Based on this summary, we appeal to the initial intuition that there exist
infinite-component fields in terms of which we can effectively describe all 
states of particles with internal degrees of freedom (the hadrons).

With regard to the theory of the ISFIR-class fields, which we consider in this 
paper, the following is immediately evident. First, the theory is 
$CPT$-invariant (for proving this following Pauli [11], it is irrelevant 
whether the sum of finite-dimensional representations of the proper Lorentz 
group is finite or infinite). Second, the (anti)commutator of any two 
components of the ISFIR-class fields is represented by a finite sum of 
derivatives of the Pauli-Jordan function $D(x-y)$, but the maximum degree of 
derivatives is unlimited on the set of all such (anti)commutators.

To outline the fundamental differences between the theory of finite-dimentional 
fields and the theory of ISFIR-class fields, we invoke a simple analogy that
gives an idea of the essence of the matter. If the physical quantities, cross 
sections first of all, can be assigned finite polynomials in some variable $x$
in the first theory, they are assigned infinite convergent series in the second
theory. If the coefficients in these series have alternating signs, then the 
quantities associated with such series can rapidly tend to zero as
$x \rightarrow +\infty$. In the two theories under discussion, the role of the 
variable $x$ belongs to the matrix elements of finite transformations for the 
irreducible representations of the proper Lorentz group; these matrix elements 
are represented by finite Laurent series in the Lorentz factor
$(1-v^{2}/c^{2})^{-1/2}$. In the theory we finally obtain as a result of our 
constraction, we really have such a series with alternating signs, in 
particular, for the differential cross section of charged lepton-proton 
scattering, if we associate the lowest-energy spin-1/2 state for one of our 
infinite-component fields with the proton (the detailed description of this 
will be the subject of another paper).

Note also that one of the most serious difficulties for the theories of 
higher-spin particles in the framework of finite-dimentional representations of 
the proper Lorentz group is the problem of renormalizability. Our knowing of 
this difficulty in no way brings us closer to solving the renormalizability 
problem for some specific theory of the ISFIR-class fields; the solution of 
this problem will be of the highest degree of sophistication by far.

Two of our paper, this one and the paper being prepared for publication, are 
devoted to a detailed description of a subclass of the theories of the 
ISFIR-class fields, which is determined by the requirement of a double symmetry 
for the Lagrangians of free and interacting fields. The double symmetry, whose
general definition is given in [12], includes relativistic invariance\footnote{
According to modern terminology and the notation in [5], relativistic 
invariance means the invariance under the orthochronous Lorentz group 
$L^{\uparrow}$ generated by the proper Lorentz group $L^{\uparrow}_{+}$ and the
spatial reflection $P$. The group $L^{\uparrow}$ is called the total Lorentz
group in [2] and [13].} (the primary symmetry) and the invariance under the 
transformations of a global secondary symmetry generated by the polar or the 
axial four-vector representation of the group $L^{\uparrow}$. These 
transformations can be written as
\begin{equation}
\Psi (x) \rightarrow \Psi '(x) = \exp [-i D^{\mu} \theta_{\mu}] \Psi (x),
\end{equation}
where the parameters $\theta_{\mu}$ are the components of a polar or an axial
four-vector of the orthochronous Lorentz group and $D^{\mu}$ are matrix 
operators.

If an element $g$  of the group $L^{\uparrow}$ is assigned the transformations
\begin{equation}
\Psi '(x) = S(g) \Psi (x),
\end{equation}
\begin{equation}
\theta'_{\mu} = [l_{\pm}(g)]_{\mu}{}^{\nu} \theta_{\nu}
\end{equation}
of the field $\Psi (x)$ and a polar or an axial four-vector $\theta_{\mu}$ (the
signs $+$ or $-$ correspond to the respective transformations of the polar or 
the axial four-vector $\theta_{\mu}$), then the operators $D^{\mu}$ must 
satisfy the relations
\begin{equation}
S^{-1}(g)D^{\nu}S(g)[l_{\pm}(g)]_{\nu}{}^{\mu} = D^{\mu}.
\end{equation}

What are the physical and mathematical motivations for invoking a secondary 
symmetry in general and the given specific symmetry in particular? First, in
addition to the primary symmetry defined by the properties of space-time, 
matter can have its own symmetry, which is related to the primary symmetry, is
consistent with it, and does not violate it. Second, using the modern ideas of 
the inner structure of the hadrons, we can interpret transformations (2) with a
polar four-vector $\theta_{\mu}$ as adding some coherent state (the 
exponential) of gluons and quark-antiquark pairs to the hadron (the field
$\Psi (x)$). In addition, we consider transformations (2) with an axial
four-vector $\theta_{\mu}$ on equal footing and restrict our consideration to 
the global transformations. Third, because all finite-dimentional irreducible
representations of the proper Lorentz group with integer spins can be obtained 
from the direct products of the four-vector representation, the restrictions
imposed by this secondary symmetry on the theory are exceeded in severity only
by the restrictions imposed by a secondary symmetry generated by the bispinor
representation of the proper Lorentz group. We have consider these latter 
restrictions in detail but do not discuss them here.

As we show in this paper, the invariance of the free-field Lagrangian under 
secondary symmetry transformations (2) along with its relativistic invariance 
provides an effective tool for selecting the proper Lorentz group 
representations and eliminating an infinite number of arbitrary constants of 
the Lagrangian. The symmetry of each Eq. (1) obtained is so high that the 
particle mass spectrum is infinitely degenerate in spin, is continuous, and 
ranges from some positive value $m_{0}$ to $+ \infty$ if $R \neq 0$. To 
eliminate this degeneracy, we assume that the secondary symmetry is 
spontaneously broken, i.e., one or several Lorentz scalar components of the 
infinite-dimentional fields under consideration have nonzero vacuum expectation
values. An analogue of this in quantum chromodynamics is the assumption of a 
quark-antiquark condensate. Therefore, before beginning a study of the 
properties of the selected class of free fields, we must first find the 
structure of the interaction Lagrangians for such fields with the required 
double symmetry. This will be the subject of another paper.

\begin{center}
{\large \bf 2. Necessary information about Lorentz group representations and
the relativistically invariant Lagrangians}
\end{center}

It seems reasonable to begin with information (taken from [2] and [13]) about 
the irreducible Lorentz group representations and the relativistically 
invariant Lagrangians that provides a basis for the calculations in this paper.
In addition, this avoids possible inconveniences due to the remote publication
of the cited sources, establishes a uniqueness of notions and notation, the 
more so, as the replacement of the Lorentz group with the group $SO(4)=SO(3) 
\otimes SO(3)$ so far prevails in describing the representations in the 
literature.

An infinitesimal proper Lorentz transformation
$$x^{\mu} \rightarrow x'^{\mu} = x^{\mu}+g^{\mu\nu}\epsilon_{\nu\rho}x^{\rho},$$
where $g^{00}=-g^{11}=-g^{22}=-g^{33}=1$, $g^{\mu\nu}=0$  ($\mu \neq \nu$), and 
$\epsilon_{\nu\rho} = -\epsilon_{\rho\nu}$, of the contravariant space-time
coordinates is assigned a transformation
$$\Psi \rightarrow \Psi' = \Psi +\frac{1}{2}\epsilon_{\mu\nu}I^{\mu\nu}\Psi,$$
where $I^{\mu\nu} =-I^{\nu\mu}$, in a representation space. The commutation
relations for the infinitesimal operators of the proper Lorentz group are
\begin{equation}
[I^{\mu\nu}, I^{\rho\sigma}] = -g^{\mu\rho}I^{\nu\sigma}+ g^{\mu\sigma}
I^{\nu\rho} + g^{\nu\rho}I^{\mu\sigma} - g^{\nu\sigma}I^{\mu\rho}.
\end{equation}

An irreducible representation $\tau$ of the proper Lorentz group is defined by 
a pair of number $(l_{0},l_{1})$, where $2l_{0}$ is an integer and $l_{1}$ is 
an arbitrary complex number. The canonical basis vectors of this representation
space are related to a subgroup $SO(3)$ and are denoted by $\xi_{\tau lm}$, 
where $l$ is a spin, $m$ is a projection of the spin onto the third axis,
$m=-l,-l+1, \ldots, l$, and $l=|l_{0}|, |l_{0}|+1, \ldots$. The representation
$\tau=(l_{0},l_{1})$ is finite-dimentional if $2l_{1}$ is an integer of the 
same parity as $2l_{0}$ and $|l_{1}| > |l_{0}|$; then $l=|l_{0}|, |l_{0}|+1, 
\ldots, |l_{1}|-1$. The pairs $(l_{0},l_{1})$ and $(-l_{0},-l_{1})$ define the
same representation, $(l_{0},l_{1}) \sim (-l_{0},-l_{1})$. If we introduce the
operators
\begin{equation}
F^{3} = iI^{30}, \hspace{0.5 cm} F^{-} = iI^{10}+I^{20}, \hspace{0.5 cm}
F^{+} = iI^{10}-I^{20},
\end{equation}
then for the irreducible representation $\tau = (l_{0}, l_{1})$, the formulas
\begin{equation}
F^{3}\xi_{\tau lm} = B_{\tau l} \sqrt{l^{2}-m^{2}} \xi_{\tau l-1 m}
- A_{\tau l} m \xi_{\tau lm}
- B_{\tau l+1} \sqrt{(l+1)^{2}-m^{2}} \xi_{\tau l+1 m},
\end{equation}
$$F^{-}\xi_{\tau lm} = -B_{\tau l} \sqrt{(l+m)(l+m-1)} 
\xi_{\tau l-1 m-1} - A_{\tau l} \sqrt{(l+m)(l-m+1)} \xi_{\tau l m-1}$$
\begin{equation}
- B_{\tau l+1} \sqrt{(l-m+1)(l-m+2)} \xi_{\tau l+1 m-1},
\end{equation}
$$F^{+}\xi_{\tau lm} = B_{\tau l} \sqrt{(l-m)(l-m-1)}
\xi_{\tau l-1 m+1} - A_{\tau l} \sqrt{(l-m)(l+m+1)} \xi_{\tau l m+1}$$
\begin{equation}
+ B_{\tau l+1} \sqrt{(l+m+1)(l+m+2)} \xi_{\tau l+1 m+1}
\end{equation}
hold, where
$$A_{\tau l} = \frac{il_{0}l_{1}}{l(l+1)}, \hspace{0.5 cm}
B_{\tau l} = \frac{i}{l}
\sqrt{\frac{(l^{2}-l_{0}^{2})(l^{2}-l_{1}^{2})}{4l^{2}-1}}.$$

In passing from the proper to the orthochronous Lorentz group, we must 
introduce the spatial-reflection operator $P$, whose action on the canonical
basis vectors is give by
$$P \xi_{\tau lm} = (-1)^{[l]}p_{\dot{\tau} \tau} \xi_{\dot{\tau} lm},$$
where $p_{\dot{\tau} \tau} p_{\tau \dot{\tau}} = 1$ and
$\dot{\tau} = (-l_{0},l_{1}) \sim (l_{0},-l_{1})$ if $\tau = (l_{0},l_{1})$.

It is conventional to indicate the representation of the proper Lorentz group,
irrespective of whether the proper or the orthochronous Lorentz group is
considered.

If a proper Lorentz group representation is decomposable into a finite or an 
infinite direct sum of finite-dimentional irreducible representations, then the
Hermitian bilinear form
$$\overline{\Psi_{2}}\Psi_{1} \equiv (\Psi_{2}, \Psi_{1})=
\sum_{\tau, l, m, \tau', l', m'} (\psi_{2})^{*}_{\tau' l'm'}
a_{\tau' l' m', \hspace{0.1 cm} \tau l m} (\psi_{1})_{\tau lm}$$
for the vectors $\Psi_{1} = \{ (\psi_{1})_{\tau lm} \}$ and $\Psi_{2} = 
\{ (\psi_{2})_{\tau lm}\}$ of this representation space is invariant under the
proper Lorentz group provided that
$$a_{\tau' l' m', \hspace{0.1 cm} \tau l m} =
(-1)^{[l]}a_{\dot{\tau} \tau} \delta_{\tau' \dot{\tau}}
\delta_{ll'} \delta_{mm'}.$$
The Hermiticity of the bilinear form implies that $a_{\dot{\tau} \tau} = 
a_{\tau \dot{\tau}}^{*}$. This form is invariant under the spatial reflection
$P$ if $a_{\dot{\tau} \tau} = p_{\dot{\tau} \tau}^{*} a_{\tau \dot{\tau}} 
p_{\dot{\tau} \tau}$.

If an element $g$ of the orthochronous Lorentz group is assigned the 
transformation
\begin{equation}
x'_{\mu} = [l_{+}(g)]_{\mu}{}^{\nu} x_{\nu}
\end{equation}
of the covariant space-time coordinates and a field transformation of form (3),
then the free Lagrangian
\begin{equation}
{\cal L}_{0} = \frac{i}{2}[(\Psi, \Gamma^{\mu} \partial_{\mu} \Psi ) - 
(\partial_{\mu} \Psi, \Gamma^{\mu} \Psi )] - \kappa (\Psi, \Psi )
\end{equation}
is relativistically invariant provided that the matrix operators $\Gamma^{\mu}$ 
satisfy the conditions
\begin{equation}
S^{-1}(g) \Gamma^{\nu} S(g) [l_{+}(g)]_{\nu}{}^{\mu} = \Gamma^{\mu}.
\end{equation}
In addition, the operator $\Gamma^{0}$ must satisfy the relation
\begin{equation}
(\Gamma^{0} \Psi_{1}, \Psi_{2}) = (\Psi_{1}, \Gamma^{0} \Psi_{2})
\end{equation}
for the Lagrangian to be real.

Its follows from Eq. (13) that
\begin{equation}
[I^{\mu\nu}, \Gamma^{\rho}] = -g^{\mu\rho} \Gamma^{\nu}
+ g^{\nu\rho} \Gamma^{\mu},
\end{equation}
\begin{equation}
[P, \Gamma^{0}] = 0,
\end{equation}
and, in particular,
\begin{equation}
[I^{i0}, \Gamma^{0}] = \Gamma^{i}, \hspace{0.5 cm} i=1,2,3.
\end{equation}

Let
\begin{equation}
\Gamma^{0} \xi_{\tau lm} = \sum_{\tau', l', m'}
c_{\tau ' l' m', \hspace{0.1 cm} \tau l m} \xi_{\tau ' l' m'}.
\end{equation}
Then conditions (15) yield
\begin{equation}
c_{\tau ' l' m', \hspace{0.1 cm} \tau l m} =
c_{\tau ' \tau} (l) \delta_{ll'} \delta_{mm'},
\end{equation}
and the quantity $c_{\tau ' \tau} (l) \equiv c(l|l'_{0},l'_{1};l_{0},l_{1})$ 
can be different from zero for a given $\tau = (l_{0},l_{1})$ only if
$\tau' = (l_{0} \pm 1,l_{1})$ or $(l_{0},l_{1} \pm 1)$. From conditions (15),
we also find
\begin{equation}
c(l|l_{0}+1,l_{1};l_{0},l_{1}) = c(l_{0}+1,l_{1};l_{0},l_{1})
\sqrt{(l+l_{0}+1)(l-l_{0})},
\end{equation}
\begin{equation}
c(l|l_{0},l_{1};l_{0}+1,l_{1}) = c(l_{0},l_{1};l_{0}+1,l_{1})
\sqrt{(l+l_{0}+1)(l-l_{0})},
\end{equation}
\begin{equation}
c(l|l_{0},l_{1}+1;l_{0},l_{1}) = c(l_{0},l_{1}+1;l_{0},l_{1})
\sqrt{(l+l_{1}+1)(l-l_{1})},
\end{equation}
\begin{equation}
c(l|l_{0},l_{1};l_{0},l_{1}+1) = c(l_{0},l_{1};l_{0},l_{1}+1)
\sqrt{(l+l_{1}+1)(l-l_{1})},
\end{equation}
where $c(l'_{0},l'_{1};l_{0},l_{1}) \equiv c_{\tau' \tau}$ are arbitrary.

Condition (16) reduces to
\begin{equation}
c_{\tau' \dot{\tau}} p_{\dot{\tau} \tau}
= p_{\tau' \dot{\tau}'} c_{\dot{\tau}' \tau},
\end{equation}
and relation (14) yields
\begin{equation}
c_{\dot{\tau} \tau'}^{*}(l) a_{\dot{\tau} \tau}=
a_{\tau' \dot{\tau}'} c_{\dot{\tau}' \tau}(l).
\end{equation}

It follows from formulas (20)-(23) that coupling of finite-dimentional and 
infinite-dimentional representations of the proper Lorentz group is impossible
in the relativistically invariant Lagrangian because $c_{\tau' \tau}(l) = 
c_{\tau \tau'} (l) = 0$ for pairs $\tau$ and $\tau'$ of this type: if $\tau = 
(\pm |l_{0}|, |l_{0}|+1)$, then there must be $\tau' = 
(\pm (|l_{0}|+1), |l_{0}|+1)$ or $\tau' = (\pm |l_{0}|, |l_{0}|)$.

If the operator $\Gamma^{0}$ is known, we use relations (17) to find the 
operators $\Gamma^{i}$, $i=1,2,3$.

\begin{center}
{\large \bf 3. Conditions imposed by the secondary symmetry on the 
infinite-component free-field theory}
\end{center}

In this paper, we find all variants of a free-field theory that satisfies the
following conditions:

{\bf Condition 1}. The proper Lorentz group representation $S$, according to 
which any of the considered fields transforms, is decomposable into an infinite
direct sum of finite-dimentional irreducible representations, and the 
multiplicity of each irreducible representation does not exceed unity. The 
bosonic fields belong to one of the two types defined by the transformation 
properties under spatial reflection: for any irreducible representation $\tau$
belonging to $S$, either $p_{\dot{\tau} \tau} = 1$ or $p_{\dot{\tau} \tau} = 
-1$.

{\bf Condition 2}. Lagrangian (12) for each field is relativistically invariant
and unsplittable, i.e., it cannot be represented as a sum of two Lagrangians 
containing no identical field components.

{\bf Condition 3}. Lagrangian (12) for each field is invariant under the 
nontrivial ($D^{\mu} \neq 0$) global transformations of secondary symmetry (2) 
generated by either the polar or the axial four-vector representation of the
orthochronous Lorentz group.

{\bf Condition 4}. Matrix operator $\Gamma^{0}$ in Lagrangian (12) is 
Hermitian, $c^{*}_{\tau' \tau}(l) = c_{\tau \tau'}(l)$.

As a basic variant in the first stage of our study, we assume (Condition 2) 
that the infinite-component bosonic free fields are described by first-order 
differential equations. It is quite possible that we must abandon this 
description in favor of second-order equations at some stage, but knowing the 
structure of the first-order equations, we can also obtain the particular class
of second-order equations without any difficulties.

It is evident that the weak-interaction Lagrangian involving the 
infinite-component fields under consideration cannot be invariant under the 
secondary symmetry transformations generated by the polar  or the axial 
four-vector representation of the group $L^{\uparrow}$. However, this 
Lagrangian, as well as the total Lagrangian, can have a well-defined secondary 
symmetry generated by a second-rank $(0,3)$ or $(-1,2) \oplus (1,2)$ tensor of 
the group $L^{\uparrow}_{+}$. We could find all variants of the theory with 
this secondary symmetry from the very beginning, but this would be a somewhat
conceptually different and, probably, more complicated problem.

Because relations (5) for the operators $D^{\mu}$ and relation (13) for the 
operators $\Gamma^{\mu}$ are the same for any elements $g$ of the proper 
Lorentz group, the matrix operators $D^{\mu}$ are described by formulas similar
to Eqs. (15) and (18)-(23). The arbitrary quantities in such formulas are 
denoted as $d_{\tau' \tau} \equiv d(l'_{0},l'_{1};l_{0},l_{1})$. The properties
of the operator $D^{0}$ under spatial reflection are described by a relation 
similar to Eq. (16) if the parameter $\theta_{\mu}$ in transformation (2) is a 
polar four-vector or by the formula $\{ P, D^{0} \} = 0$ if $\theta_{\mu}$ is 
an axial four-vector.

Lagrangian (12) is invariant under secondary symmetry transformations (2) if 
the relations
\begin{equation}
[\Gamma^{\mu}, D^{\nu}] = 0,
\end{equation}
\begin{equation}
(D^{0} \Psi_{1}, \Psi_{2}) = (\Psi_{1}, D^{0} \Psi_{2})
\end{equation}
hold. If the constant $\kappa$ in Lagrangian (12) is equal to zero, the 
invariance of the Lagrangian under secondary symmetry transformations (2) is
provided by either conditions (26) and (27) or the conditions
\begin{equation}
\{ \Gamma^{\mu}, D^{\nu} \} = 0,
\end{equation}
\begin{equation}
(D^{0} \Psi_{1}, \Psi_{2}) = -(\Psi_{1}, D^{0} \Psi_{2}).
\end{equation}

Define the matrix operator $\Gamma^{5}$ by
$$\Gamma^{5} \xi_{(l_{0},l_{1})lm}=(-1)^{l_{0}+l_{1}} \xi_{(l_{0},l_{1})lm}.$$
It is easy to verify that any four-vector operator $V^{\mu}$ (satisfying a
condition of type (5)) anticommutes with the operator $\Gamma^{5}$. 
Consequently, if the operators $V^{\mu}$ and $W^{\nu}$ commute with each other,
then the operators $V^{\mu}$ and $\Gamma^{5} W^{\nu}$ anticommute; this implies
that if we know the solutions of Eqs. (26), we also know the solutions of Eqs. 
(28), and vice versa. In what follows, we therefore deal only with Eqs. (26).

Take two independent relations
\begin{equation}
[\Gamma^{0},D^{0}]=0,
\end{equation}
\begin{equation}
[\Gamma^{0},D^{3}]=0
\end{equation}
from conditions (26). The remaining Eqs. (26) follow from formulas (30) and 
(31) and commutation relations like (15) and (6).

Using the relation $D^{3}=[I^{30}, D^{0}]$ and formulas (7), (8), and (18)-(23),
we reduce  relations (30) and (31) to an algebraic system of equations with 
respect to the quantities $c_{\tau' \tau}$ and $d_{\tau' \tau}$. Note that Eqs.
(32)-(41) are valid for any completely reducible Lorentz group representation
$S$ consistent with Condition 2 and with Condition 1 only in part, namely, no
two irreducible representations of the group $L^{\uparrow}_{+}$ belonging to 
$S$ are equivalent, while "boundary" equations (42)-(44) hold for the 
representations $S$ that, in addition, satisfy one more part of Condition 1, 
i.e., the irreducible representations belonging to $S$ are finite-dimentional.
 
In the case where the representation $(l_{0}, l_{1})$ contains no less than two
values of spin ($|l_{1}| \geq |l_{0}|+2$), we have
\begin{equation}
c(l_{0}+1,l_{1};l_{0},l_{1}) d(l_{0},l_{1};l_{0}-1,l_{1}) -
d(l_{0}+1,l_{1};l_{0},l_{1}) c(l_{0},l_{1};l_{0}-1,l_{1}) =0,
\end{equation}
\begin{equation}
c(l_{0},l_{1}+1;l_{0},l_{1}) d(l_{0},l_{1};l_{0},l_{1}-1) -
d(l_{0},l_{1}+1;l_{0},l_{1}) c(l_{0},l_{1};l_{0},l_{1}-1) =0,
\end{equation}
\vspace{-0.2 cm}
$$l_{1}c(l_{0}+1,l_{1}+1;l_{0}+1,l_{1}) d(l_{0}+1,l_{1};l_{0},l_{1}) 
-(l_{0}+1)d(l_{0}+1,l_{1}+1;l_{0}+1,l_{1})c(l_{0}+1,l_{1};l_{0},l_{1})$$ 
$$+l_{0}c(l_{0}+1,l_{1}+1;l_{0},l_{1}+1) d(l_{0},l_{1}+1;l_{0},l_{1})$$
\begin{equation}
-(l_{1}+1)d(l_{0}+1,l_{1}+1;l_{0},l_{1}+1)c(l_{0},l_{1}+1;l_{0},l_{1})=0,
\end{equation}
\vspace{-0.2 cm}
$$c(l_{0}+1,l_{1}+1;l_{0}+1,l_{1}) d(l_{0}+1,l_{1};l_{0},l_{1}) -
d(l_{0}+1,l_{1}+1;l_{0}+1,l_{1})c(l_{0}+1,l_{1};l_{0},l_{1})$$
$$+c(l_{0}+1,l_{1}+1;l_{0},l_{1}+1) d(l_{0},l_{1}+1;l_{0},l_{1})$$
\begin{equation}
-d(l_{0}+1,l_{1}+1;l_{0},l_{1}+1)c(l_{0},l_{1}+1;l_{0},l_{1})=0,
\end{equation}
\vspace{-0.2 cm}
$$l_{1}c(l_{0},l_{1}+1;l_{0},l_{1}) d(l_{0},l_{1};l_{0}+1,l_{1}) +
l_{0}d(l_{0},l_{1}+1;l_{0},l_{1})c(l_{0},l_{1};l_{0}+1,l_{1})$$
$$-(l_{0}+1)c(l_{0},l_{1}+1;l_{0}+1,l_{1}+1)d(l_{0}+1,l_{1}+1;l_{0}+1,l_{1})$$
\begin{equation}
-(l_{1}+1)d(l_{0},l_{1}+1;l_{0}+1,l_{1}+1)c(l_{0}+1,l_{1}+1;l_{0}+1,l_{1})=0,
\end{equation}
\vspace{-0.2 cm}
$$c(l_{0},l_{1}+1;l_{0},l_{1}) d(l_{0},l_{1};l_{0}+1,l_{1}) -
d(l_{0},l_{1}+1;l_{0},l_{1})c(l_{0},l_{1};l_{0}+1,l_{1})$$
$$+c(l_{0},l_{1}+1;l_{0}+1,l_{1}+1) d(l_{0}+1,l_{1}+1;l_{0}+1,l_{1})$$
\begin{equation}
-d(l_{0},l_{1}+1;l_{0}+1,l_{1}+1)c(l_{0}+1,l_{1}+1;l_{0}+1,l_{1})=0,
\end{equation}
\vspace{-0.2 cm}
$$c(l_{0},l_{1};l_{0}-1,l_{1}) d(l_{0}-1,l_{1};l_{0},l_{1}) -
d(l_{0},l_{1};l_{0}-1,l_{1}) c(l_{0}-1,l_{1};l_{0},l_{1})$$ 
$$+c(l_{0},l_{1};l_{0}+1,l_{1}) d(l_{0}+1,l_{1};l_{0},l_{1}) -
d(l_{0},l_{1};l_{0}+1,l_{1}) c(l_{0}+1,l_{1};l_{0},l_{1})$$ 
$$+c(l_{0},l_{1};l_{0},l_{1}-1) d(l_{0},l_{1}-1;l_{0},l_{1}) -
d(l_{0},l_{1};l_{0},l_{1}-1) c(l_{0},l_{1}-1;l_{0},l_{1})$$
\begin{equation}
+c(l_{0},l_{1};l_{0},l_{1}+1) d(l_{0},l_{1}+1;l_{0},l_{1}) -
d(l_{0},l_{1};l_{0},l_{1}+1) c(l_{0},l_{1}+1;l_{0},l_{1}) =0,
\end{equation}
\vspace{-0.2 cm}
$$(l_{0}-1)[c(l_{0},l_{1};l_{0}-1,l_{1})d(l_{0}-1,l_{1};l_{0},l_{1})+
d(l_{0},l_{1};l_{0}-1,l_{1}) c(l_{0}-1,l_{1};l_{0},l_{1})]$$
$$-(l_{0}+1)[c(l_{0},l_{1};l_{0}+1,l_{1}) d(l_{0}+1,l_{1};l_{0},l_{1})+
d(l_{0},l_{1};l_{0}+1,l_{1}) c(l_{0}+1,l_{1};l_{0},l_{1})]$$
$$+(l_{1}-1)[c(l_{0},l_{1};l_{0},l_{1}-1) d(l_{0},l_{1}-1;l_{0},l_{1})+
d(l_{0},l_{1};l_{0},l_{1}-1) c(l_{0},l_{1}-1;l_{0},l_{1})]$$
\begin{equation}
-(l_{1}+1)[c(l_{0},l_{1};l_{0},l_{1}+1)d(l_{0},l_{1}+1;l_{0},l_{1})+
d(l_{0},l_{1};l_{0},l_{1}+1)c(l_{0},l_{1}+1;l_{0},l_{1})]=0,
\end{equation}
\vspace{-0.2 cm}
$$l_{1}[c(l_{0},l_{1};l_{0}-1,l_{1})d(l_{0}-1,l_{1};l_{0},l_{1})+
d(l_{0},l_{1};l_{0}-1,l_{1}) c(l_{0}-1,l_{1};l_{0},l_{1})$$
$$-c(l_{0},l_{1};l_{0}+1,l_{1}) d(l_{0}+1,l_{1};l_{0},l_{1})-
d(l_{0},l_{1};l_{0}+1,l_{1}) c(l_{0}+1,l_{1};l_{0},l_{1})]$$
$$+l_{0}[c(l_{0},l_{1};l_{0},l_{1}-1) d(l_{0},l_{1}-1;l_{0},l_{1})+
d(l_{0},l_{1};l_{0},l_{1}-1) c(l_{0},l_{1}-1;l_{0},l_{1})$$
\begin{equation}
-c(l_{0},l_{1};l_{0},l_{1}+1)d(l_{0},l_{1}+1;l_{0},l_{1})-
d(l_{0},l_{1};l_{0},l_{1}+1)c(l_{0},l_{1}+1;l_{0},l_{1})]=0,
\end{equation}
\vspace{-0.2 cm}
$$l_{0}(l_{0}-1)[c(l_{0},l_{1};l_{0}-1,l_{1})d(l_{0}-1,l_{1};l_{0},l_{1})-
d(l_{0},l_{1};l_{0}-1,l_{1})c(l_{0}-1,l_{1};l_{0},l_{1})]$$
$$+l_{0}(l_{0}+1)[c(l_{0},l_{1};l_{0}+1,l_{1}) d(l_{0}+1,l_{1};l_{0},l_{1})-
d(l_{0},l_{1};l_{0}+1,l_{1}) c(l_{0}+1,l_{1};l_{0},l_{1})]$$
$$+l_{1}(l_{1}-1)[c(l_{0},l_{1};l_{0},l_{1}-1)d(l_{0},l_{1}-1;l_{0},l_{1})-
d(l_{0},l_{1};l_{0},l_{1}-1)c(l_{0},l_{1}-1;l_{0},l_{1})]$$
\begin{equation}
+l_{1}(l_{1}+1)[c(l_{0},l_{1};l_{0},l_{1}+1) d(l_{0},l_{1}+1;l_{0},l_{1})-
d(l_{0},l_{1};l_{0},l_{1}+1) c(l_{0},l_{1}+1;l_{0},l_{1})]=0,
\end{equation}
and the quantities $c_{\tau' \tau}$ and $d_{\tau' \tau}$ that could describe
coupling of the finite-dimentional and infinite-dimentional irreducible 
representations of the group $L^{\uparrow}_{+}$ must be omitted in accordance
with the previous remark.

In addition to relations (32)-(41), there are also six equations\footnote{In
referring to these six equations, we assign them the same numbers, but with
primes.} obtained from Eqs. (32)-(37) with the replacement $c_{\tau' \tau} 
\rightarrow c_{\tau \tau'}$, $d_{\tau' \tau} \rightarrow d_{\tau \tau'}$. 

If $l_{1}=|l_{0}|+1$, the quantities $c_{\tau' \tau}$ and $d_{\tau' \tau}$ must
satisfy the "boundary" equations
$$c(l_{0} \pm 1,|l_{0}|+2;l_{0},|l_{0}|+2)d(l_{0},|l_{0}|+2;l_{0},|l_{0}|+1)$$
\vspace{-0.7 cm}
\begin{equation}
+ d(l_{0} \pm 1,|l_{0}|+2;l_{0},|l_{0}|+2) 
c(l_{0},|l_{0}|+2;l_{0},|l_{0}|+1) = 0,
\end{equation}
\vspace{-0.2 cm}
$$c(l_{0},|l_{0}|+1;l_{0} \mp 1,|l_{0}|+1) 
d(l_{0} \mp 1,|l_{0}|+1;l_{0},|l_{0}|+1)$$
$$+ d(l_{0},|l_{0}|+1;l_{0} \mp 1,|l_{0}|+1) 
c(l_{0} \mp 1,|l_{0}|+1;l_{0},|l_{0}|+1)$$
$$+ c(l_{0},|l_{0}|+1;l_{0},|l_{0}|+2) d(l_{0},|l_{0}|+2;l_{0},|l_{0}|+1)$$
\begin{equation}
+ d(l_{0},|l_{0}|+1;l_{0},|l_{0}|+2) c(l_{0},|l_{0}|+2;l_{0},|l_{0}|+1) = 0,
\end{equation}
\vspace{-0.2 cm}
$$l_{0}[c(l_{0},|l_{0}|+1;l_{0} \mp 1,|l_{0}|+1)
d(l_{0} \mp 1,|l_{0}|+1;l_{0},|l_{0}|+1)$$
$$- d(l_{0},|l_{0}|+1;l_{0} \mp 1,|l_{0}|+1)
c(l_{0} \mp 1,|l_{0}|+1;l_{0},|l_{0}|+1)]$$
$$- (l_{0}+1)[c(l_{0},|l_{0}|+1;l_{0},|l_{0}|+2) 
d(l_{0},|l_{0}|+2;l_{0},|l_{0}|+1)$$
\begin{equation}
- d(l_{0},|l_{0}|+1;l_{0},|l_{0}|+2) c(l_{0},|l_{0}|+2;l_{0},|l_{0}|+1)] = 0,
\end{equation}
where the upper and lower signs correspond to the respective cases $l_{0} 
\geq 0$ and $l_{0} \leq 0$. One more equation must be added to these equations;
it is obtained from Eq. (42) with the replacement $c_{\tau' \tau} \rightarrow 
c_{\tau \tau'}$ and $d_{\tau' \tau} \rightarrow d_{\tau \tau'}$.

It is easy to verify that the quantities $c_{\tau' \tau}$ and $d_{\tau' \tau}$,
which are written later in Corollaries 1-4, satisfy all Eqs. (32)-(44) and also
provide the necessary transformation properties of the operators $\Gamma^{\mu}$ 
and $D^{\mu}$ under spatial reflection and satisfy relations (14) and (27). The
proof that Corollaries 1-4 exhaust all variants of the theory satisfying 
Conditions 1-4 is too cumbersome; we therefore restrict ourselves to outlining 
it and present some general conclusions at the end of this section and some 
specific conclusions at the beginning of the next section, which allow 
reconstructing the full body of the proof.

A simple analysis of Eqs. (32)-(37) and (42) shows that the equality of all the
quantities $c_{\tau \tau'}d_{\tau' \tau}$ to zero is only possible if either 
field Lagrangian (12) is splittable or the secondary symmetry transformations 
are trivial, which respectively contradicts Condition 2 or Condition 3. 

Because of the equivalence of the irreducible representations $(l_{0},l_{1})$ 
and $(-l_{0}, -l_{1})$, we can restrict ourselves to only pairs $(l_{0},l_{1})$
such that $l_{1} > 0$.

For the proper Lorentz group representation \ $S$ \ satisfying Condition 1, let 
there exist a solution of system (32)-(44) that is consistent with Conditions 2
and 3. Then among the representations $(l_{0},l_{1}) \in S$, there is a 
representation $(k_{0},k_{1})$ such that either 
$c(k_{0},k_{1}+1;k_{0},k_{1}) d(k_{0},k_{1};k_{0},k_{1}+1) \neq 0$ or
$d(k_{0},k_{1}+1;k_{0},k_{1}) c(k_{0},k_{1};k_{0},k_{1}+1) \neq 0$ and the
quantities $c(l_{0},l_{1}+1;l_{0},l_{1}) d(l_{0},l_{1};l_{0},l_{1}+1)$ and
$d(l_{0},l_{1}+1;l_{0},l_{1}) c(l_{0},l_{1};l_{0},l_{1}+1)$ are equal to zero
provided that $|l_{0}| \leq l_{1}-1$ and $l_{1} \leq k_{1}-1$. If 
$k_{0} \neq 0$, it follows from Eqs. (38)-(41), (43), and (44) that
\begin{equation}
d(l_{0}+1,k_{1};l_{0},k_{1}) c(l_{0},k_{1};l_{0}+1,k_{1}) =
\frac{(k_{1}+k_{0})a-(k_{1}-k_{0})b}{2k_{1}(k_{1}-l_{0})(k_{1}+l_{0}+1)},
\end{equation}
\begin{equation}
c(l_{0}+1,k_{1};l_{0},k_{1}) d(l_{0},k_{1};l_{0}+1,k_{1}) =
\frac{-(k_{1}-k_{0})a+(k_{1}+k_{0})b}{2k_{1}(k_{1}-l_{0})(k_{1}+l_{0}+1)},
\end{equation}
\begin{equation}
d(l_{0},k_{1}+1;l_{0},k_{1}) c(l_{0},k_{1};l_{0},k_{1}+1) =
\frac{-(l_{0}+k_{0})a+(l_{0}-k_{0})b}
{(k_{1}-l_{0})(k_{1}-l_{0}+1)(k_{1}+l_{0})(k_{1}+l_{0}+1)},
\end{equation}
\begin{equation}
c(l_{0},k_{1}+1;l_{0},k_{1}) d(l_{0},k_{1};l_{0},k_{1}+1) =
\frac{(l_{0}-k_{0})a-(l_{0}+k_{0})b}
{(k_{1}-l_{0})(k_{1}-l_{0}+1)(k_{1}+l_{0})(k_{1}+l_{0}+1)},
\end{equation}
where at least one of the constants $a$ or $b$ is nonzero, $-k_{1}+1 \leq l_{0}
\leq k_{1}-2$ in Eqs. (45) and (46), and $|l_{0}| \leq k_{1}-1$ in Eqs. (47) 
and (48). If $k_{0} = 0$, the numerators in the right-hand sides of Eqs. 
(45)-(48) must be respectively replaced with $k_{1}(A-B)$, $k_{1}(A+B)$, 
$-k_{1}A+l_{0}B$, and $-k_{1}A-l_{0}B$, where at least one of the constants $A$ 
or $B$ is nonzero.

To proceed further in analyzing the system of equations (32)-(44), we introduce
the quantities $g_{\tau' \tau}$ such that
\begin{equation}
d_{\tau' \tau} = g_{\tau' \tau} c_{\tau' \tau}
\end{equation}
if $c_{\tau' \tau} \neq 0$. Having verified that the last inequality holds, we
can use the following equations obtained from the respective 
equations\footnote{See footnote 3.} (42), (32), (33), (34) and (35),
(36) and (37), (34) and ($34'$), and (36) and ($36'$):
\begin{equation}
g(l_{0},l_{0}+2;l_{0},l_{0}+1) = -g(l_{0}+1,l_{0}+2;l_{0},l_{0}+2),
\end{equation}
\vspace{-0.4 cm}
\begin{equation}
g(l_{0},l_{1};l_{0}-1,l_{1}) = g(l_{0}+1,l_{1};l_{0},l_{1}),
\end{equation}
\vspace{-0.2 cm}
\begin{equation}
g(l_{0},l_{1};l_{0},l_{1}-1) = g(l_{0},l_{1}+1;l_{0},l_{1}),
\end{equation}
\vspace{-0.2 cm}
$$[g(l_{0}+1,l_{1}+1;l_{0},l_{1}+1) - g(l_{0},l_{1}+1;l_{0},l_{1})]$$
$$\times [l_{1}g(l_{0}+1,l_{1};l_{0},l_{1}) -
(l_{0}+1)g(l_{0}+1,l_{1}+1;l_{0}+1,l_{1})]$$
$$=[g(l_{0}+1,l_{1};l_{0},l_{1}) - g(l_{0}+1,l_{1}+1;l_{0}+1,l_{1})]$$
\begin{equation}
\times [(l_{1}+1)g(l_{0}+1,l_{1}+1;l_{0},l_{1}+1) -
l_{0}g(l_{0},l_{1}+1;l_{0},l_{1})],
\end{equation}
\vspace{-0.2 cm}
$$[g(l_{0},l_{1}+1;l_{0}+1,l_{1}+1) - g(l_{0}+1,l_{1}+1;l_{0}+1,l_{1})]$$
$$\times [l_{1}g(l_{0},l_{1};l_{0}+1,l_{1}) + 
l_{0}g(l_{0},l_{1}+1;l_{0},l_{1})]$$
$$=[g(l_{0},l_{1};l_{0}+1,l_{1}) - g(l_{0},l_{1}+1;l_{0},l_{1})]$$
\begin{equation}
\times [(l_{1}+1)g(l_{0},l_{1}+1;l_{0}+1,l_{1}+1) +
(l_{0}+1)g(l_{0}+1,l_{1}+1;l_{0}+1,l_{1})],
\end{equation}
\vspace{-0.2 cm}
$$g(l_{0}+1,l_{1}+1;l_{0},l_{1}+1) g(l_{0},l_{1}+1;l_{0},l_{1})$$
$$\times [l_{1}g(l_{0}+1,l_{1};l_{0},l_{1}) -
(l_{0}+1)g(l_{0}+1,l_{1}+1;l_{0}+1,l_{1})]$$
$$\times [l_{1}g(l_{0},l_{1};l_{0}+1,l_{1}) -
(l_{0}+1)g(l_{0}+1,l_{1};l_{0}+1,l_{1}+1)]$$
$$\times c(l_{0},l_{1};l_{0}+1,l_{1}) d(l_{0}+1,l_{1};l_{0},l_{1})
c(l_{0}+1,l_{1};l_{0}+1,l_{1}+1) d(l_{0}+1,l_{1}+1;l_{0}+1,l_{1})$$
$$=g(l_{0}+1,l_{1}+1;l_{0}+1,l_{1}) g(l_{0}+1,l_{1};l_{0},l_{1})$$
$$\times [(l_{1}+1)g(l_{0}+1,l_{1}+1;l_{0},l_{1}+1) -
l_{0}g(l_{0},l_{1}+1;l_{0},l_{1})]$$
$$\times [(l_{1}+1)g(l_{0},l_{1}+1;l_{0}+1,l_{1}+1) -
l_{0}g(l_{0},l_{1};l_{0},l_{1}+1)]$$
\begin{equation}
\times c(l_{0},l_{1};l_{0},l_{1}+1) d(l_{0},l_{1}+1;l_{0},l_{1})
c(l_{0},l_{1}+1;l_{0}+1,l_{1}+1) d(l_{0}+1,l_{1}+1;l_{0},l_{1}+1),
\end{equation}
\vspace{-0.2 cm}
$$g(l_{0}+1,l_{1}+1;l_{0}+1,l_{1}) g(l_{0}+1,l_{1}+1;l_{0},l_{1}+1)$$
$$\times [l_{1}g(l_{0},l_{1};l_{0}+1,l_{1}) + l_{0}g(l_{0},l_{1}+1;l_{0},l_{1})]
[l_{1}g(l_{0}+1,l_{1};l_{0},l_{1}) + l_{0}g(l_{0},l_{1};l_{0},l_{1}+1)]$$
$$\times c(l_{0},l_{1};l_{0}+1,l_{1}) d(l_{0}+1,l_{1};l_{0},l_{1})
c(l_{0},l_{1};l_{0},l_{1}+1) d(l_{0},l_{1}+1;l_{0},l_{1})$$
$$=g(l_{0}+1,l_{1};l_{0},l_{1}) g(l_{0},l_{1}+1;l_{0},l_{1})$$
$$\times [(l_{1}+1)g(l_{0},l_{1}+1;l_{0}+1,l_{1}+1) +
(l_{0}+1)g(l_{0}+1,l_{1}+1;l_{0}+1,l_{1})]$$
$$\times [(l_{1}+1)g(l_{0}+1,l_{1}+1;l_{0},l_{1}+1) +
(l_{0}+1)g(l_{0}+1,l_{1};l_{0}+1,l_{1}+1)]$$
$$\times c(l_{0}+1,l_{1};l_{0}+1,l_{1}+1)d(l_{0}+1,l_{1}+1;l_{0}+1,l_{1})$$
\begin{equation}
\times c(l_{0},l_{1}+1;l_{0}+1,l_{1}+1) d(l_{0}+1,l_{1}+1;l_{0},l_{1}+1),
\end{equation}
and also the equations obtained from Eqs. (50)-(54) with the replacement 
$g_{\tau' \tau} \rightarrow g_{\tau \tau'}$.

\begin{center}
{\large \bf 4. Consequences of the double symmetry for the fermionic free-field
Lagrangian}
\end{center}

Whatever the quantities $p_{\dot{\tau} \tau}$ and $a_{\dot{\tau} \tau}$ may be
in some canonical basis, we can find a new canonical basis such that formulas 
(8)-(10) remain unchanged and the equalities
\begin{equation}
p_{\dot{\tau} \tau} = p_{\tau \dot{\tau}} = 1,
\end{equation}
\begin{equation}
a_{\dot{\tau} \tau} = a_{\tau \dot{\tau}} = \pm 1
\end{equation}
hold; the signs in the right-hand side of Eq. (58) can be either identical or 
opposite for the different representations $\tau$.

Using relations (24) and (25) and Condition 4 and taking formulas (20)-(23) 
into account, we obtain
\begin{equation}
c(l_{0}+1,l_{1};l_{0},l_{1}) = c(-l_{0}-1,l_{1};-l_{0},l_{1}),
\end{equation}
\begin{equation}
c(l_{0},l_{1} \pm 1;l_{0},l_{1}) = c(-l_{0},l_{1} \pm 1;-l_{0},l_{1}),
\end{equation}
\begin{equation}
c(l_{0}+1,l_{1};l_{0},l_{1})=c^{*}(l_{0},l_{1};l_{0}+1,l_{1}),
\end{equation}
\begin{equation}
c(l_{0},l_{1}+1;l_{0},l_{1})=-c^{*}(l_{0},l_{1};l_{0},l_{1}+1),
\end{equation}
and also find that the quantities $a_{\dot{\tau} \tau}$ for all the 
representations $\tau$ are the same and can be set equal to unity without loss 
of generality. Note that the last statement and relations (61) and (62) fail if
we omit Condition 4.

Consider the case where the parameter $\theta_{\mu}$ in transformations (2) is 
a polar four-vector. The quantities $d_{\tau' \tau}$ must then satisfy the 
conditions of type (59) and (60). It follows from formulas (59) for
$c_{\tau' \tau}$ and $d_{\tau' \tau}$ and from (45) and (46) for $l_{0} = -1/2$ 
that $a = b$, after which it follows from Eqs. (45)-(48) and (60) that
 \begin{equation}
g(l_{0}+1,k_{1};l_{0},k_{1}) = g(l_{0},k_{1};l_{0}+1,k_{1}),
\end{equation}
\begin{equation}
g(l_{0},k_{1}+1;l_{0},k_{1}) = g(l_{0},k_{1};l_{0},k_{1}+1),
\end{equation}
\begin{equation}
g(l_{0},k_{1}+1;l_{0},k_{1}) = g(-l_{0},k_{1}+1;-l_{0},k_{1})
\end{equation}
for all admissible values of $l_{0}$.

We must analyze two possible cases,
\begin{equation} 
c(k_{1}, k_{1}+1; k_{1}-1, k_{1}+1)d(k_{1}-1, k_{1}+1; k_{1}, k_{1}+1) \neq 0,
\end{equation}
or
\begin{equation}
c(k_{1}, k_{1}+1; k_{1}-1, k_{1}+1)d(k_{1}-1, k_{1}+1; k_{1}, k_{1}+1) = 0,
\end{equation}
separately.

We dwell on the first case. It follows from Eqs. (42) and (32) that we can 
introduce the quantities $g(l_{0} \pm 1, k_{1}+1; l_{0}, k_{1}+1)$ for all 
admissible values of $l_{0}$. First using equality (50) for $l_{0}=k_{1}-1$ and
equality (51) for $l_{1}=k_{1}+1$ and all admissible values of $l_{0}$, we find
that the system of equations (53) and (54) for $l_{0}=k_{1}-2$ and 
$l_{1}=k_{1}$ has two solutions; we then catch a dependence of the quantities
$g(l_{0}, k_{1}+1; l_{0}, k_{1})$ on $l_{0}$ for each of the solutions and 
verify this dependence by induction. We find that one of the solutions of the 
system of equations (53) and (54) is
\begin{equation}
g(l_{0},k_{1}+1;l_{0},k_{1})= (-1)^{l_{0}+k_{1}-1} g_{0},
\end{equation}
where $g_{0}$ is a constant. This solution contradicts relation (65) and must 
be rejected. Only the solution
$$g(l_{0} \pm 1,k_{1};l_{0},k_{1})=g(l_{0} \pm 1,k_{1}+1;l_{0},k_{1}+1)$$
\begin{equation}
=-g(l_{0},k_{1}+1;l_{0},k_{1})=-g(l_{0},k_{1};l_{0},k_{1}+1)=g_{0}
\end{equation}
remains. Using relations (45)-(48) (for $a=b$) and (69), we now turn to the 
system of equations (55) and (56) for $l_{0}=k_{1}-2$ and $l_{1}=k_{1}$. 
Because $a \neq 0$ and $g_{0} \neq 0$, this system is consistent if and only if
$k_{1} = 3/2$. For such a value of $k_{1}$, we have 
$$c(\frac{1}{2}, \frac{5}{2}; -\frac{1}{2}, \frac{5}{2})
d(-\frac{1}{2}, \frac{5}{2}; \frac{1}{2}, \frac{5}{2})
= d(\frac{1}{2}, \frac{5}{2}; -\frac{1}{2}, \frac{5}{2})
c(-\frac{1}{2}, \frac{5}{2}; \frac{1}{2}, \frac{5}{2}) = \frac{ak_{0}}{6}.$$
Thereafter, we can again use the system of equations (38)-(41), (43), and (44)
for $l_{1}=k_{1}+1=5/2$, and so on. Finally, using induction, we can verify that
the representation S incorporates all finite-dimentional irreducible 
half-integer spin representations of the group $L^{\uparrow}_{+}$ and that for 
all admissible values $l_{0}$ and $l_{1}$, the equalities
\begin{equation}
g(l_{0} \pm 1,l_{1};l_{0},l_{1}) = -g(l_{0},l_{1}\pm 1;l_{0},l_{1}) = g_{0},
\end{equation}
\begin{equation}
c(l_{0},l_{1};l_{0}+1,l_{1})c(l_{0}+1,l_{1};l_{0},l_{1}) =
c(l_{0},l_{1};l_{0},l_{1}+1)c(l_{0},l_{1}+1;l_{0},l_{1}) = f_{0}
\end{equation}
hold, where $f_{0}$ is a constant. It is evident that solution (71) contradicts 
relations (61) and (62) related to Condition 4, i.e., Conditions 1-4 cannot be
fulfiled in the case under consideration. If we give up Condition 4, then taking
formulas (58), (25),(20)-(23), and (71) into account, we find that Conditions 
1-3 are fulfiled if and only if
$$a_{\dot{\tau} \tau} = (-1)^{l_{1}-3/2} e_{0} \hspace{0.5 cm} 
\mbox{{\rm for}} \hspace{0.5 cm} \tau = (l_{0}, l_{1}),$$
$$c(l_{0}+1,l_{1};l_{0},l_{1}) = c(l_{0},l_{1};l_{0}+1,l_{1})
= c(l_{0},l_{1}+1;l_{0},l_{1}) = c(l_{0},l_{1};l_{0},l_{1}+1) = c_{0},$$
where $c_{0}$ is a real constant and the constant $e_{0}$ equals either 1 or -1.
Here and in what follows, the values of the quantities $c_{\tau' \tau}$ are 
presented up to inessential phase factors that can be introduced or eliminated 
by changing the relative phases of the basis vectors of the space of the 
representation $S$.

We do not analyze the case described by equality (67) nor any other variants 
here.

To simplify the formulations of Corollaries 1-4, we first note that each 
representation $S$ of the group $L^{\uparrow}_{+}$ in these corollaries is 
assigned a unique set of the quantities $c_{\tau' \tau}$ and $d_{\tau' \tau}$
(up to a common normalization factor), which can be written for all appropriate
pairs $\tau$ and $\tau'$ belonging to $S$, and a unique set of the quantities
$a_{\dot{\tau} \tau}$ (up to a common sign fixed to be positive), 
$a_{\dot{\tau} \tau} = 1$ for all $\tau \in S$.

{\bf Corollary 1}. {\it The requirement that a fermionic free-field theory 
satisfy Conditions} 1-4 {\it if the parameter} $\theta_{\mu}$ {\it in 
transformations} (2) {\it is a polar four-vector can be fulfiled only for a 
countable set of the proper Lorentz group representations that are numbered by 
half-integer numbers} $k_{1}$ 
($k_{1} \geq 3/2$),
\begin{equation}
S^{k_{1}} = \sum^{+\infty}_{n_{1}=0} \sum^{k_{1}-3/2}_{n_{0}=-k_{1}+1/2}
\oplus (\frac{1}{2}+n_{0}, k_{1}+n_{1}),
\end{equation}
{\it and the quantities} $c_{\tau' \tau}$ {\it and} $d_{\tau' \tau}$ {\it 
corresponding to the representation} $S^{k_{1}}$ {\it are given by the 
equalities}
\begin{equation}
c(l_{0}+1,l_{1};l_{0},l_{1}) =c(l_{0},l_{1};l_{0}+1,l_{1})
=c_{0} \sqrt{\frac{(k_{1}-l_{0}-1)(k_{1}+l_{0})}
{(l_{1}-l_{0})(l_{1}-l_{0}-1) (l_{1}+l_{0}) (l_{1}+l_{0}+1)}},
\end{equation}
\begin{equation}
c(l_{0},l_{1}+1;l_{0},l_{1}) =c(l_{0},l_{1};l_{0},l_{1}+1)
=c_{0} \sqrt{\frac{(k_{1}-l_{1}-1)(k_{1}+l_{1})}
{(l_{1}-l_{0})(l_{1}-l_{0}+1) (l_{1}+l_{0}) (l_{1}+l_{0}+1)}},
\end{equation}
\begin{equation}
d(l_{0}+1,l_{1};l_{0},l_{1}) =d(l_{0},l_{1};l_{0}+1,l_{1}) =
g_{0} c(l_{0}+1,l_{1};l_{0},l_{1}),
\end{equation}
\begin{equation}
d(l_{0},l_{1}+1;l_{0},l_{1}) =d(l_{0},l_{1};l_{0},l_{1}+1) =
g_{0} c(l_{0},l_{1}+1;l_{0},l_{1}),
\end{equation}
{\it where} $c_{0}$ {\it and} $g_{0}$ {\it are real constants}. 

{\bf Corollary 2}. {\it The requirement that a fermionic free-field theory 
satisfy Conditions} 1-4 {\it if the parameter} $\theta_{\mu}$ {\it in 
transformations} (2) {\it is an axial four-vector can be fulfiled only in the 
following three cases}:

1. {\it for the countable set of the proper Lorentz group representations whose
element is given by formula} (72), {\it where} $k_{1} \geq 3/2$, {\it with the
quantities} $c_{\tau' \tau}$ {\it corresponding to representation} $S^{k_{1}}$ 
{\it given by formulas} (73) {\it and} (74), {\it where} $c_{0}$ {\it is a real
constant, and the quantities} $d_{\tau' \tau}$ {\it given by}
\begin{equation}
d(l_{0}+1,l_{1};l_{0},l_{1}) = -d(l_{0},l_{1};l_{0}+1,l_{1}) =
g_{0}l_{1} c(l_{0}+1,l_{1};l_{0},l_{1}),
\end{equation}
\begin{equation}
d(l_{0},l_{1}+1;l_{0},l_{1}) = -d(l_{0},l_{1};l_{0},l_{1}+1) =
g_{0}l_{0} c(l_{0},l_{1}+1;l_{0},l_{1}),
\end{equation}
{\it where} $g_{0}$ {\it is a real constant};

2. {\it for the countable set of the proper Lorentz group representations whose
element is given by formula} (72), {\it where} $k_{1} \geq 3/2$, {\it with the
quantities} $c_{\tau' \tau}$ {\it and} $d_{\tau' \tau}$ {\it corresponding to 
representation} $S^{k_{1}}$ {\it given by}
$$c(l_{0}+1,l_{1};l_{0},l_{1}) =c(l_{0},l_{1};l_{0}+1,l_{1})=$$
\begin{equation}
=(-1)^{l_{1}-1/2} c_{0}l_{1} \sqrt{\frac{(k_{1}-l_{0}-1)(k_{1}+l_{0})}
{(l_{1}-l_{0})(l_{1}-l_{0}-1) (l_{1}+l_{0}) (l_{1}+l_{0}+1)}},
\end{equation}
$$c(l_{0},l_{1}+1;l_{0},l_{1}) =c(l_{0},l_{1};l_{0},l_{1}+1)=$$
\begin{equation}
=(-1)^{l_{0}-1/2} c_{0}l_{0} \sqrt{\frac{(k_{1}-l_{1}-1)(k_{1}+l_{1})}
{(l_{1}-l_{0})(l_{1}-l_{0}+1) (l_{1}+l_{0}) (l_{1}+l_{0}+1)}},
\end{equation}
\begin{equation}
d(l_{0}+1,l_{1};l_{0},l_{1}) =-d(l_{0},l_{1};l_{0}+1,l_{1}) =
g_{0}l_{1}^{-1} c(l_{0}+1,l_{1};l_{0},l_{1}),
\end{equation}
\begin{equation}
d(l_{0},l_{1}+1;l_{0},l_{1}) =-d(l_{0},l_{1};l_{0},l_{1}+1) =
g_{0}l_{0}^{-1} c(l_{0},l_{1}+1;l_{0},l_{1}),
\end{equation}
{\it where} $c_{0}$ {\it and} $g_{0}$ {\it are real constants; and}

3. {\it for the proper Lorentz group representation} $S^{F}$ {\it containing all
finite-dimentional irreducible half-integer spin representations of the group}
$L^{\uparrow}_{+}$,
\begin{equation}
S^{F} = \sum^{+\infty}_{n_{1}=0} \sum^{n_{1}}_{n_{0}=-n_{1}-1}
\oplus (1/2+n_{0}, 3/2+n_{1}),
\end{equation}
{\it with the corresponding quantities} $c_{\tau' \tau}$ {\it and} 
$d_{\tau' \tau}$ {\it given by}
$$c(l_{0}+1,l_{1};l_{0},l_{1}) = c(l_{0},l_{1};l_{0}+1,l_{1})=$$
\begin{equation}
=(-1)^{l_{1}+1/2} c_{0} \sqrt{\frac{1 - (-1)^{l_{1}+l_{0}}}
{2(l_{1}-l_{0}-1)(l_{1}+l_{0})} + \frac{1 + (-1)^{l_{1}+l_{0}}}
{2(l_{1}+l_{0}+1)(l_{1}-l_{0})}},
\end{equation}
$$c(l_{0},l_{1}+1;l_{0},l_{1}) = c(l_{0},l_{1};l_{0},l_{1}+1)=$$
\begin{equation}
=c_{0} \sqrt{\frac{1 - (-1)^{l_{1}+l_{0}}}{2(l_{0}-l_{1}-1)(l_{1}+l_{0})} 
+ \frac{1 + (-1)^{l_{1}+l_{0}}}{2(l_{1}+l_{0}+1)(l_{0}-l_{1})}},
\end{equation}
\begin{equation}
d(l_{0}+1,l_{1};l_{0},l_{1}) = -d(l_{0},l_{1};l_{0}+1,l_{1}) =
(-1)^{l_{1}+1/2} g_{0} c(l_{0}+1,l_{1};l_{0},l_{1}),
\end{equation}
\begin{equation}
d(l_{0},l_{1}+1;l_{0},l_{1}) = -d(l_{0},l_{1};l_{0},l_{1}+1) =
(-1)^{l_{0}-1/2} g_{0} c(l_{0},l_{1}+1;l_{0},l_{1}),
\end{equation}
{\it where} $c_{0}$ {\it and} $g_{0}$ {\it are real constants}.

\begin{center}
{\large \bf 5. Consequences of the double symmetry for the bosonic free-field
Lagrangian}
\end{center}

Corollaries 3 and 4 given below are valid for either of the two types of bosonic
described in Condition 1.

{\bf Corollary 3}. {\it The requirement that a bosonic free-field theory satisfy 
Conditions} 1-4 {\it if the parameter} $\theta_{\mu}$ {\it in transformations} 
(2) {\it is a polal four-vector can be fulfiled only in the following two cases}:

1. {\it for a countable set of the proper Lorentz group representations numbered
by integer numbers} $k_{1}$ ($k_{1} \geq 1$),
\begin{equation}
S^{k_{1}} = \sum^{+\infty}_{n_{1}=0} \sum^{k_{1}-1}_{n_{0}=-k_{1}+1}
\oplus (n_{0}, k_{1}+n_{1}),
\end{equation}
{\it with the corresponding quantities} $c_{\tau' \tau}$ {\it and} 
$d_{\tau' \tau}$ {\it respectively given by formulas} (73) {\it and} (74) {\it 
and formulas} (75) {\it and} (76), {\it where} $c_{0}$ {\it and} $g_{0}$ {\it 
are real constants};

2. {\it for the proper Lorentz group representation containing all
finite-dimentional irreducible integer-spin representations of the group}
$L^{\uparrow}_{+}$, {\it which is denoted by} $S^{B}$,
\begin{equation}
S^{B} = \sum^{+\infty}_{n_{1}=0} \sum^{n_{1}}_{n_{0}=-n_{1}}
\oplus (n_{0}, 1+n_{1}),
\end{equation}
{\it with the corresponding quantities} $c_{\tau' \tau}$ {\it and} 
$d_{\tau' \tau}$ {\it given by}
$$c(l_{0}+1,l_{1};l_{0},l_{1}) = c(l_{0},l_{1};l_{0}+1,l_{1})=$$
\begin{equation}
=(-1)^{l_{1}+1} c_{0}\sqrt{\frac{1 + (-1)^{l_{1}+l_{0}}}
{2(l_{1}+l_{0}+1)(l_{1}-l_{0}-1)} + \frac{1 - (-1)^{l_{1}+l_{0}}}
{2(l_{1}+l_{0})(l_{1}-l_{0})}},
\end{equation}
$$c(l_{0},l_{1}+1;l_{0},l_{1}) = c(l_{0},l_{1};l_{0},l_{1}+1)=$$
\begin{equation}
=c_{0} \sqrt{\frac{1 + (-1)^{l_{1}+l_{0}}}{2(l_{1}+l_{0}+1)(l_{0}-l_{1}-1)}
+\frac{1 - (-1)^{l_{1}+l_{0}}}{2(l_{1}+l_{0})(l_{0}-l_{1})}},
\end{equation}
\begin{equation}
d(l_{0}+1,l_{1};l_{0},l_{1}) = d(l_{0},l_{1};l_{0}+1,l_{1}) =
(-1)^{l_{1}+1} g_{0} c(l_{0}+1,l_{1};l_{0},l_{1}),
\end{equation}
\begin{equation}
d(l_{0},l_{1}+1;l_{0},l_{1}) = d(l_{0},l_{1};l_{0},l_{1}+1) =
(-1)^{l_{0}} g_{0} c(l_{0},l_{1}+1;l_{0},l_{1}),
\end{equation}
{\it where} $c_{0}$ {\it and} $g_{0}$ {\it are real constants}.

{\bf Corollary 4}. {\it The requirement that a bosonic free-field theory satisfy 
Conditions} 1-4 {\it if the parameter} $\theta_{\mu}$ {\it in transformations} 
(2) {\it is an axial four-vector can be fulfiled only for the countable set of 
the proper Lorentz group representations whose element is given by formula} 
(88), {\it where} $k_{1} \geq 2$, {\it with the quantities} $c_{\tau' \tau}$ 
{\it and} $d_{\tau' \tau}$ {\it corresponding to representation} $S^{k_{1}}$ 
{\it respectively given by formulas} (73) {\it and} (74) {\it and formulas} (77)
{\it and} (78), {\it where} $c_{0}$ {\it and} $g_{0}$ {\it are real constants}.

It seems reasonable to also incorporate the variants of the bosonic field 
theory, where the following condition holds either together with or instead of 
Condition 3.

{\bf Condition 3A}. Both types of bosonic free fields, denoted by $\varphi_{+}$ 
and $\varphi_{-}$, are described by identical Lagrangians before a spontaneous 
breaking of the secondary symmetry. The sum of these Lagrangians is invariant
under the secondary symmetry transformations given by
\begin{equation}
\left(
\begin{array}{c}
\varphi_{+} \\ \varphi_{-}
\end{array}
\right) \rightarrow \left(
\begin{array}{c}
\varphi'_{+} \\ \varphi'_{-}
\end{array}
\right) = \exp \left[ -i \left(
\begin{array}{lr}
0            & D^{\mu} \\
D^{\mu}  & 0
\end{array}
\right) \theta_{\mu} \right] \left(
\begin{array}{c}
\varphi_{+} \\ \varphi_{-}
\end{array}
\right) , 
\end{equation}
where the parameter $\theta_{\mu}$ are the components of a polar or an axial 
four-vector of the orthochronous Lorentz group and $D^{\mu}$ are matrix 
operators.

It is evident that if the parameter $\theta_{\mu}$ is a polar (axial) 
four-vector, then the bosonic free-field theory satisfies Condition 1, 2, 3A, 
and 4 for those representation $S$ and their corresponding quantities
$c_{\tau' \tau}$ and $d_{\tau' \tau}$ that are described in Corollary 4 
(Corollary 3).

\begin{center}
{\large \bf 6. Concluding remarks}
\end{center}

In each variant of the theory described in Corollaries 1-4, the operators
$D^{\mu}$ in secondary symmetry transformations (2) are defined uniquely up to a
common normalization constant; therefore, the secondary symmetry group generated 
by transformations (2) and their products is also defined uniquely. In the 
present paper, we have no need to reveal the characteristics of the Lie algebra 
for any of these groups. We only note that the Lie algebra of the double 
symmetry group corresponding to Corollaries 1, 2 (part 2), and 3 (part 1) 
coincides with the Lie algebra of the Poincar\'{e} group; however, the operators
$D^{\mu}$ cannot be identified with the translation generators $P^{\mu}$, 
because the $D^{\mu}$ act on the spin variables of the field, while the 
$P^{\mu}$ do not.

Lagrangian (12) is assigned Gelfand-Yaglom equation (1), where $R = \kappa E$
and $E$ is an identity operator. Let $\lambda$ be an eigenvalue of the operators
$\Gamma^{0}$,
$$\Gamma^{0} \Psi (\lambda ) = \lambda \Psi (\lambda ).$$
Then the mass $M$ spectrum obtained from Eq. (1) is related to the eigenvalue
$\lambda$ spectrum for the operator $\Gamma^{0}$ by $M =\kappa /\lambda$. 
Because of the secondary symmetry of the theory, the operator $\Gamma^{0}$ 
commutes with the operator $D^{3}$, which, according to an equality similar to 
Eq. (17) and relations (7) and (8), takes a state with certain spin $l$ to the 
states with spins $l-1$, $l$, and  $l+1$; therefore, the vectors $(D^{3})^{n} 
\Psi (\lambda )$, $n \geq 1$, are also the eigenvectors with the eigenvalue
$\lambda$ of the operator $\Gamma^{0}$. The mass spectrum corresponding to each 
infinite-component field of the theory under consideration is thus infinitely 
degenerate in spin. A detailed analysis shows that the mass spectrum is 
continuous, discrete points are absent from the spectrum. However, the infinite 
degeneracy in spin is already sufficient for posing a question of a spontaneous 
breaking of the secondary symmetry.

{\bf Acknowledgments}. The author is very grateful to A.U. Klimyk, A.A. Komar,
and V. I. Fushchich for the useful discussions of the problems considered in 
this paper.


\begin{thebibliography}{99}

\bibitem{1}
   V.L.Ginzburg and I.E.Tamm, {\it Zh.Eksp.Teor.Fiz.} {\bf 17} (1947) 227.
\bibitem{2}
   I.M.Gelfand and A.M.Yaglom, {\it Zh.Eksp.Teor.Fiz.} {\bf 18} (1948) 703.
\bibitem{3}
   A.A.Komar and L.M.Slad, {\it Teor.Mat.Fiz.} {\bf 1} (1969) 50.
\bibitem{4}
   I.T.Grodsky and R.F.Streater, {\it Phys.Rev.Lett.} {\bf 20} (1968) 695.
\bibitem{5}
   N.N.Bogolubov, A.A.Logunov, A.I.Oksak, and I.T.Todorov, 
   {\it General principles of 
   quantum field theory} (Kluwer Academic Publishers, Dordrecht, 1990).
\bibitem{6}
   I.M.Gelfand and A.M.Yaglom, {\it Zh.Eksp.Teor.Fiz.} {\bf 18} (1948) 1094.
\bibitem{7}
   V.Bargmann, {\it Math.Rev.} {\bf 10} (1949) 583 and 584.   
\bibitem{8}
   E.Abers, I.T.Grodsky, and R.E.Norton, {\it Phys.Rev.} {\bf 159} (1967) 1222. 
\bibitem{9}
   P.A.M.Dirac, {\it Proc.Roy.Soc.} {\bf A322} (1971) 435.
\bibitem{10}
   L.M.Slad, {\it Teor.Mat.Fiz.} {\bf 2} (1970) 67.
\bibitem{11}
   W.Pauli, in: {\it Niels Bohr and the development of physics} (Edited by 
   W.Pauli with assistance of L.Rosenfeld and V.Weisskopf, Pergamon Press, 
   London, 1955), p.30.
\bibitem{12}
   L.M.Slad, {\it Mod.Phys.Lett.} {\bf A15} (2000) 379 (hep-th/0003107).
\bibitem{13}
   I.M.Gelfand, R.A.Minlos and Z.Ya.Shapiro, 
   {\it Representations of the roration and Lorenz 
   group and their applications} (The Macmillan Company, New York, 1963).

\end{thebibliography}
\end{document}